\documentclass[letterpaper]{article}

\usepackage{aaai24}
\usepackage{times}
\usepackage{helvet}
\usepackage{courier}
\usepackage{tikz}
\usepackage{cite}
\usepackage{soul}
\usepackage{xcolor}
\usepackage{subfig}
\usepackage{amsmath}
\usepackage{physics}
\usepackage{booktabs}
\usepackage{amsfonts}
\usepackage{graphicx}
\usepackage{textcomp}
\usepackage{algorithmic}
\usepackage{filecontents}
\usepackage[font=small]{caption}
\usepackage{esvect}
\usepackage{natbib}
\usepackage[linesnumbered,lined,boxed,commentsnumbered]{algorithm2e}
\usepackage[format=plain,labelfont={bf,it},labelsep=period,textfont=it]{caption}
\usepackage{numprint}
\npthousandsep{,}            

\usepackage{tcolorbox,tikz}
\tcbuselibrary{skins}
\tcbuselibrary{breakable}

\newtcolorbox{mybox}[3][]
{
  breakable, 
  enhanced,
  colback  = #2!5, 
  colframe= #2!50,
  boxsep=-0.5mm,
  borderline west={1.5mm}{0.05mm}{#3!30}, 
  #1,
}

\definecolor{blue_color}{rgb}{0.0, 0.0, 1.0}
\definecolor{red_color}{rgb}{1.0, 0.0, 0.0}

\newif\ifdraft
\drafttrue

\ifdraft
  \newcommand{\todo}[1]{{\textcolor{blue_color}{TODO: #1 }}}
  \newcommand{\attn}[1]{{\textcolor{red}{DT: #1 }}}
\else
  \newcommand{\todo}[1]{}
  \newcommand{\attn}[1]{}
\fi

\newcommand{\sol}{OrganiQ}

\title{OrganiQ: Mitigating Classical Resource Bottlenecks of Quantum Generative Adversarial Networks on NISQ-Era Machines}
\author {
    Daniel Silver,\textsuperscript{\rm 1}
    Tirthak Patel,\textsuperscript{\rm 2}
    Aditya Ranjan,\textsuperscript{\rm 1}
    William Cutler,\textsuperscript{\rm 1}
    Devesh Tiwari\textsuperscript{\rm 1}
}
  
\affiliations {
    \textsuperscript{\rm 1}Northeastern University\\
    \textsuperscript{\rm 2}Rice University\\
    \{silver.da, ranjan.ad, cutler.wi, d.tiwari\}@northeastern.edu, patel.ti@rice.edu
}

\date{June 2023}

\begin{document}

\maketitle

\begin{abstract}






Driven by swift progress in hardware capabilities, quantum machine learning has emerged as a research area of interest. Recently, quantum image generation has produced promising results. However, prior quantum image generation techniques rely on classical neural networks, limiting their quantum potential and image quality. To overcome this, we introduce \sol{}, the first quantum GAN capable of producing high-quality images without using classical neural networks.


\end{abstract}

\section{Introduction}
\label{sec:intro}
Quantum machine learning has seen rapid development, with advancements spawning the field of quantum image generation \cite{tsang2023hybrid, florian2022quantum}. Quantum image generation has been developing in part due to potential speedups over similar classical algorithms \cite{lloyd2018quantum,tian2023recent, kao2023exploring}.

Quantum generative adversarial networks (GAN) have been particularly of interest for quantum image generation due to the classical success of GANs \cite{silver2023mosaiq}. Quantum GANs are essentially quantum equivalents of classical generative adversarial networks. GANs consist of a competing generative network known as a generator and a rival network to assess the produced quality of the generator, known as the discriminator. Classical GANs are widely used, generating high-quality data spanning domains such as image generation, audio generation, and generation of novel pharmaceuticals \cite{karras2018progressive, marafioti2019adversarial, blanchard2021using}. Motivated by classical success, Quantum GANs are being increasingly explored for quantum image generation with early promising results~\cite{dallaire2018quantum,huang2021experimental,hu2019quantum, tsang2023hybrid}.

In particular, prior efforts have demonstrated the generation of human-recognizable hand-written digits in simulation and on real machines, albeit as expected,  while achieving only limited image quality. However, we discover that these encouraging results come with significant limitations due to fundamental design choices in prior Quantum GAN works~\cite{dallaire2018quantum,huang2021experimental,hu2019quantum, tsang2023hybrid}. Below, we describe these limitations and the scope for enhancement.

\subsection*{Limitations of Prior Quantum GANs}

The fundamental design limitation of prior Quantum GANs is the use of a hybrid structure -- quantum in some parts, and classical in others. A representative and specific example is, QCPatch, which leverages a patch-based hybrid model, where multiple quantum generators combine to form an image which is then passed through a classical discriminator, and uses a quantum generator to generate images from quantum noise~\cite{hu2019quantum}. While a reasonable early design with promising results as a classical discriminator is useful to achieve stability in training, this approach suffers from a fundamental bottleneck -- running into classical resource bottleneck because the power of a quantum discriminator is ultimately ``held back'' by the classical discriminator during training. 

A second limitation is that a hybrid GAN design cannot natively process quantum data on a classical discriminator, and hence, the potential for hybrid QGANs are limited \cite{Lloyd_2018}. Finally, the hybrid network structure creates imbalance issues which can be difficult to tune and mitigate \cite{zoufal2019quantum}. Driven by mitigating the limitations of hybrid networks, we introduce \sol{}, a Quantum GAN consisting of a quantum generator and a quantum discriminator.

\subsection*{Contributions of \sol{}}

\noindent \textbf{I.} To the best of our knowledge, \sol{}~\footnote{\sol{} is an or\underline{\textit{gan}}ically grown solution for quantum GANs without using hybrid fertilizer from the classical neural networks.} is the first image-scale quantum GAN that is both effective and free from classical resource bottlenecks -- because it proposes to use a quantum generator and a quantum discriminator and is inherently suited for processing quantum data. 

\noindent \textbf{II.} \sol{}'s design demonstrates how the a use of a novel amplitude encoding technique can enable the generation of high-quality images for a quantum GAN by providing a method to reliably decode amplitudes. \sol{} also provides a novel technique, Unitary Injection, to improve image generation quality on quantum GANs by improving how data is expressed on the circuit.

\noindent \textbf{III.} Our experiments in simulation and using a real quantum machine  (IBM Lagos) demonstrate that \sol{} produces higher-quality images than state-of-the-art models without the potential resource bottlenecks posed by classical parts. We evaluate \sol{} on the  MNIST and Fashion MNIST datasets on IBM's Lagos quantum computer where we observe significant visual improvements over the state-of-the-art, and over 100 points improvement using the Fréchet Inception Distance (FID), a commonly used metric for image quality. Our extensive evaluation demonstrates the benefits of different key components of \sol{} and its robustness across different datasets. 

Upon acceptance, the code and resources for \sol{} will be made publicly available.





\section{Background and Related Work}

In this section, we present the relevant background for \sol{} in addition to the limitations of the state-of-the-art QGANs.

\vspace{2mm}

\noindent\textbf{The qubit} is the building block of quantum computing, analogous to the classical bit, but providing advantages by leveraging quantum properties. While a bit stores a piece of information that is binary, represented as a  $0$ or a $1$, qubits can store information in a probabilistic state between $0$ and $1$ known as a superposition. Measuring a qubit collapses the quantum state, $\psi$ to one of two basis states, $0$ or $1$ probabilistically with the probability of measuring $0$ as $\abs{\alpha}^2$ and measuring $1$ as $\abs{\beta}^2$ where $\abs{\alpha}^2 + \abs{\beta}^2 = 1$, where $\alpha$ and $\beta$ can be complex \cite{nielsen2001quantum}. Quantum logic gates are unitary operations, $U$, which can act on multiple qubits at once to create an interdependent system in a process known as entanglement, where none of the qubits can be described independently of each other. While qubits can be very powerful, the current NISQ-era of quantum computing is defined by small machines with high levels of noise \cite{patel2023toward}. \\

\noindent\textbf{Variational quantum circuits} (VQCs) are a type of parameterized quantum circuits that have been widely used for NISQ-era quantum machine learning tasks including classification, natural language processing, and reinforcement learning \cite{silver2022quilt,patel2022optic, ranjan2024proximl, silver2023sliq, di2022dawn, chen2020variational}. VQCs are a system comprised of fixed gates $X$ and tunable gates $\theta$ such that $\psi = U(X, \theta)$. Objective functions can be setup on the VQCs whose minima correspond to the desired solution. Similar to classical neural networks, we can calculate  gradients with respect to a specified loss function, which help in tuning the gates to find the optimal solution. VQCs are functionally similar to neural networks, however, they present unique challenges. As VQCs become larger, they become increasingly difficult to simulate on classical computers \cite{chen2020variational}, making training difficult. Additionally, while VQCs are one of the most likely candidates for providing quantum advantage, they possess many well known issues such as poor accuracy and training inefficiency \cite{cerezo2021variational}.

\subsection{Generative Adversarial Networks}
Generative Adversarial Networks (GANs) are neural network systems designed to transform Gaussian noise into samples that model a given distribution of data \cite{goodfellow2014generative}. The general architecture of a GAN is composed of two components, a generator responsible for synthesizing the new data and a discriminator which effectively trains the generator by learning the difference between the real data and synthesized data. The workflow is to simultaneously train these components in an adversarial game.
This game continues until an equilibrium is reached where the generator and discriminator no longer improve. This minimax game is encapsulated as  
\begin{equation}
    \min _G \max _D \mathbb{E}_{x}[\log D(\boldsymbol{x})] + \mathbb{E}_{z}[\log (1-D(G(\boldsymbol{z})))]
\end{equation}

Where $G$ represents the generator, $D$ represents the discriminator, $\boldsymbol{x}$ is the ``real data'' and $\boldsymbol{z}$ is the noise distribution that is passed into the generator. 
\subsection{Related Work and its Limitations}

\begin{figure}[t]
    \centering
    \includegraphics[scale=0.5]{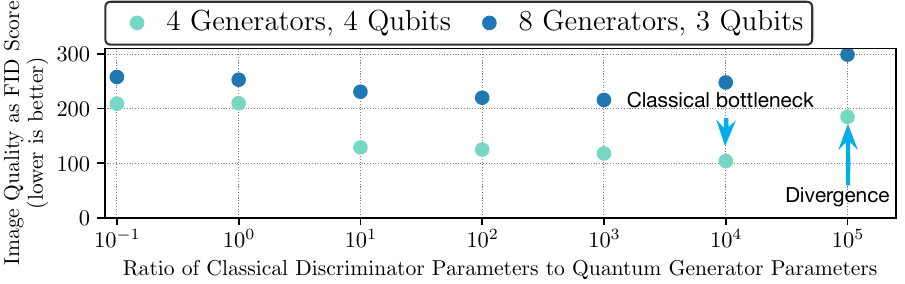}
        \vspace{-1mm}

    \caption{Image quality on the MNIST digit 3 after 50 iterations of training using the QCPatch technique. Higher ratio indicates classical discriminator needs to have more parameters than quantum generator to match quantum generator’s computing capability.}
    \label{fig:Motivation}
\end{figure}

There has been considerable research in quantum GANs, due to proposed theoretical advantages over classical GANs \cite{lloyd2018quantum, hu2019quantum,zoufal2019quantum, dallaire2018quantum}. Most of related research into quantum GANs has focused on small-scale implementations of theoretical quantum GANs that perform simple tasks quite well. While these designs provide theoretical designs and, in some cases, small implementations, they do not provide an implementation on the scale of image generation, which sets \sol{} apart.

The most relevant technique to \sol{} is a patch-based quantum GAN, comprised of quantum generators and a single classical generation which we refer to as QCPatch \cite{PhysRevApplied.16.024051}. This technique is capable of generating human-recognizable images, demonstrating effectiveness on the MNIST dataset. Typically, these classical discriminators provide stability to an otherwise, hard-to-train quantum model. However, while this model provides high-quality images, it is ultimately limited by the classical discriminator. Given a sufficiently powerful quantum generator, the classical discriminator will not be able to keep up in an adversarial game.  In Figure \ref{fig:Motivation} we demonstrate this effect as we compare the efficacy of adversarial training with a quantum generator and classical discriminator of varying size. We find the training of the generator and discriminator produces the highest quality image of parameters tested, when the classical discriminator is 10,000 times as large as the quantum generator in one architecture and 1,000 in another architecture. The different optimal points point to a non trivial tuning procedure required to balance quantum and classical components optimally.  We observe that the discriminator needs to scale significantly to compete optimally with a quantum generator, forming a bottleneck. We also note that selecting the ideal size for a classical discriminator likely requires tuning to develop a successful adversarial scheme with a given quantum generator given a critical point is reached where the quantum generator and discriminator discriminator diverge and quality worsens.

\sol{} is distinct from these methods as it uses a quantum discriminator instead of a classical discriminator. Additionally, these networks are not compatible with processing quantum data, where the generator can natively process quantum data, the classical discriminator will not have this native capability and therefore not be ideal for this task.

The goal of \sol{} is to introduce the new generation of quantum GANs that comprise a quantum generator and a quantum discriminator in order to fully take advantage of quantum properties that underline the promises of advantages quantum GANs can provide.
\section{Design Elements}
\label{sec:design}
In this section, we discuss the design and implementation of \sol{} and the procedure for training and inference.

\vspace{2mm}

\begin{figure}[t]
    \centering
    \includegraphics[scale=0.24]{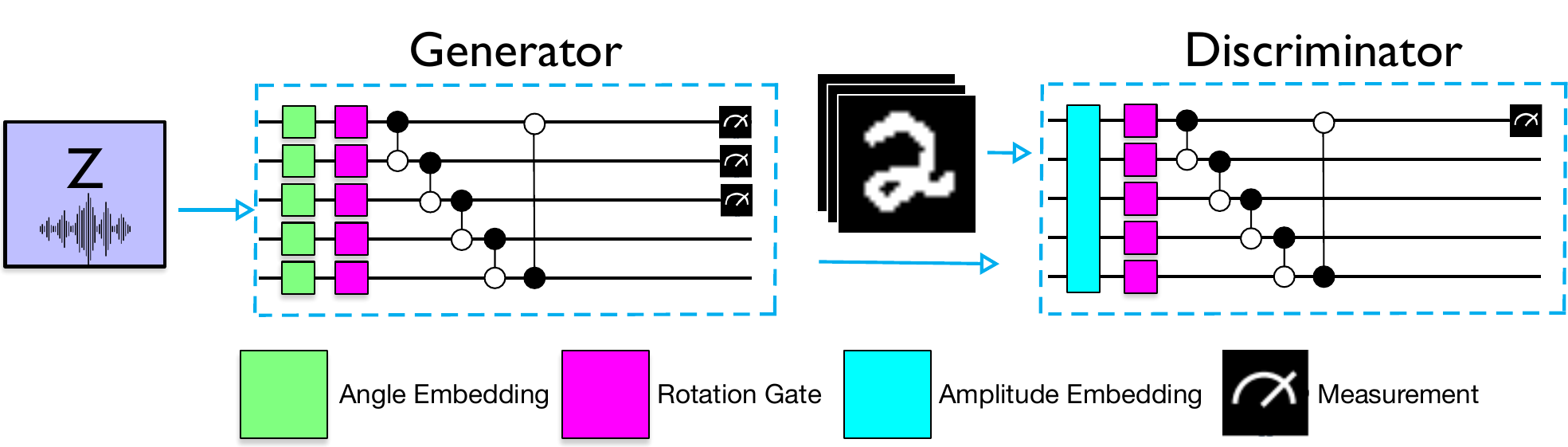}
    \caption{The Baseline is a GAN architecture ported onto quantum machines. The quantum generator produces an image that is measured classically and tested on the quantum discriminator with amplitude embedding. The discriminator also receives real (non-generated) images separately, which it learns to differentiate from the generated images.}
    \label{fig:Baseline}
    \vspace{-3mm}
\end{figure}

\noindent\textbf{Baseline Design.}
We propose a baseline design, which \sol{} is built up from. The Baseline consists of an implementation using a Quantum Generator and a Quantum Discriminator of equal size. The Baseline trains the generator and discriminator simultaneously. We use PCA to scale to full sized images with fewer qubits. The Baseline serves to demonstrate what a simpler implementation of a quantum GAN (both discriminator and generator being quantum) would be and the relative inefficacy of using such an approach. Additionally, since none of the prior works have built a quantum image GAN composed of a quantum generator and a quantum discriminator, we include the Baseline as a reasonable comparison point for \sol{}. We show a graphic of the Baseline in Figure \ref{fig:Baseline}. 

\vspace{2mm}

\noindent\textbf{Mitigating Challenges of Hybrid Quantum GANs}
Recall, hybrid quantum GANs comprised of a quantum generator and a classical discriminator have several shortcomings with both scalability and trainability. A classical discriminator adds a scalability bottleneck in training as it will not be able to keep up with a sufficiently large quantum generator. Additionally, a classical discriminator is not natively compatible with quantum data. For example, quantum state tomography reconstruction and quantum noise spectroscopy domains require quantum data \cite{perrier2022qdataset}. For these reasons, we devise a GAN consisting of purely quantum components. In order to maintain a quantum state between the generator and discriminator, as opposed to measuring in between classically, we combine the networks during training in a connected configuration.

\subsection{Combining the Generator and Discriminator}
A key design element of \sol{} is the placement of the generator and discriminator on the same quantum circuit. While previous image-based Quantum GAN approaches have typically had an architecture containing a quantum generator and a classical discriminator~\cite{PhysRevApplied.16.024051}, \sol{} proposes a new approach that leverages the quantum properties more organically by making both generator and discriminator quantum. Keeping the discriminator and generator on the same circuit facilitates smooth training where the quantum state from the generator is unperturbed throughout the training pipeline.  If the network was not connected (as is the case with prior hybrid designs), the quantum state would collapse between the generator and discriminator, causing a phase collapse to the real axis, potentially removing information stored in the complex plane. 

\vspace{2mm}

\begin{figure*}[t]
    \centering
    \includegraphics[scale=0.3]{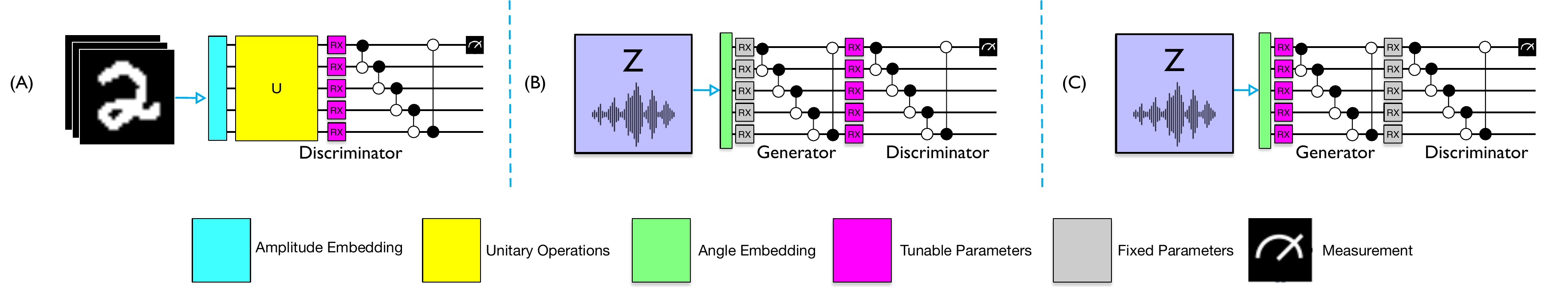}
    \caption{This figure demonstrates the three different passes as part of \sol{}'s training procedure. In \textbf{A}, the discriminator is trained on real images from the initial dataset, where amplitude embedding places the classical data on the quantum circuit. This is followed by a randomized unitary matrix which introduces phases to the classical image. Then, the discriminator is trained to update the tunable RX gates through gradient descent. In \textbf{B}, the discriminator is trained on synthetic data by placing the generator on the discriminator circuit. The generator weights are greyed out to indicate the gradients do not optimize the generator. In \textbf{C}, the generator is trained, where the discriminator's weights are greyed out to indicate that only the generator's weights are optimized in this step.}
    \label{fig:Passes}
    \vspace{-2mm}
\end{figure*}

\noindent\textbf{Training the Discriminator on Real Images} is the first step in training. This step infers the Real Loss $\text{Loss}_R$ from the discriminator $D$ output of the real images $R$. In place of the generator, we use the amplitude embedding mechanism to encode the data on the discriminator. Amplitude embedding is an encoding strategy that places $2^n$ classical features on $n$ qubits by encoding the features into real-valued amplitudes, maintaining the constraint in quantum systems that the sum of all amplitudes squared must sum to 1~\cite{Schuld_2019}. Following amplitude embedding, the remaining elements of the circuit are the basic entanglement gates and tunable weights that comprise the discriminator. We show this pass of Real data in A of Figure \ref{fig:Passes}. Note, as we use binary cross entropy as our loss function, which requires non-negative labels; in our implementation, we normalize the discriminator output which ranges from $[-1,1]$, to $[0,1]$. 

\vspace{2mm}

\noindent\textbf{Training the Discriminator on Generated Images}
In the case of training the discriminator on the generated images, we maintain the same discriminator weights and entanglements from the previous training step. The difference between this training step and the previous step lies with the image embedding portion of the circuit. In this case, instead of using amplitude embedding to place the images on the circuit, the generator circuit is used instead in order to train the discriminator on the generator's outputs. This involves passing the noise onto the generator using angle embedding, using the weights of the generator, and directly interpreting the output on the discriminator circuit. The noise is a randomized tensor uniformly distributed from $[0, 2\pi]$ of size (batch\_size, $n$), where $n$ is the number of qubits on the circuit. The circuit transforms the noise into the discriminator's classification decision. In order to only train the discriminator in this step, the generator's weights are excluded from the gradient descent procedure in training and therefore left out of optimization. After training the discriminator on fake images, the total discriminator loss $L_{D}$ is calculated by summing the Real Loss $L_{R}$ and the Fake Loss $L_{F}$.

\vspace{2mm}

\noindent\textbf{Training the Generator}
The circuits are the same here as in training the discriminator on generated images. The only difference between how these are trained is the weights in the gradient descent procedure. Instead of omitting the generator weights during gradient descent, we omit the discriminator weights during gradient descent in order to isolate and train the generator. The generator loss function uses the same binary cross-entropy function as the discriminator, however, the objective here is reversed, where the generator aims to maximize the discriminator's loss and fool the discriminator. While we have a way to train the generator to provide data to the discriminator, we need a method to be able to decode the information from the generator in inference. We therefore introduce amplitude regularization as a solution for decoding the generated quantum information to an interpretable classical space.


\subsection{Amplitude Regularization}

In the hybrid implementation of quantum GANs, encoding of the real images is normalized before being amplitude embedded and passed through the discriminator. For example, given a feature vector $[1/{\sqrt{2}}, 0, 1/{\sqrt{2}}, 0]$, the state of the 2-qubit quantum circuit would be $1/{\sqrt{2}}\ket{00} + 0\ket{01} + 1/{\sqrt{2}}\ket{10} + 0\ket{11}$. In this case, the data has been normalized, so we can directly obtain the encoded features by measuring the probability of each computational basis state and  taking their square root. However, most real-world data is not in a quantum-normalized form. Consider another case where we would like to encode  $[1/{\sqrt{2}}, 1/{\sqrt{2}}, 1/{\sqrt{2}}, 0]$ on a quantum circuit. The encoded amplitudes must be normalized and therefore becomes $1/{\sqrt{3}}\ket{00} + 1/{\sqrt{3}}\ket{00}  + 1/{\sqrt{3}}\ket{10} + 0\ket{11}$. Note that the feature vectors differ only in one value, but the encoding of other features within the vector has changed as well. Moreover, the data can not be retrieved in the exact form as reversing becomes intractable. This encoding inconsistency which also leads to a decoding inconsistency, impedes the training of the GAN and hampers the generation of images from the output of the generator.

To mitigate this inconsistency, we introduce an amplitude regularization technique. Amplitude regularization is designed to provide a uniform encoding for each feature independent of the values of other features. To do this, we reserve one state whose amplitude works as a regularization variable.  Given $n$ qubits, we normalize $2^n-1$ features $f_{i}$ as:
\begin{equation}
    \ket{\psi} = \sum_{i=0}^{2^n-1}\dfrac{1}{2^n}f_{i}\ket{i} + r\ket{2^n}
\end{equation}
where $|\psi\rangle$ represents the quantum state after the embedding and $r$ represents the amplitude of the reserved state, which we denote here as $\ket{2^n}$. Now, the probability of all the states must add up to one, which is mathematically given by
\begin{equation}
    \sum_{i=0}^{2^n-1}\big{|}\big{|}{\dfrac{1}{2^n}f_i}\big{|}\big{|}^2+||r||^2 = 1
\end{equation}
From this equation, one can obtain the value of the regularization value and encode the features into the circuit.
This ensures that any feature $f_i$ in features $f$ will be encoded invariably regardless of the other features $f_j$ where $j\neq i$. 

Although the regularization technique uniformizes the encoding scheme, it limits the space encoded states can occupy. Note that these limitations are only applicable while encoding a large amount of classical data over a small number of qubits. The effect of these limitations can be minimized by allocating more reserved states for regularization and amplitude embedding over partitions of the larger circuit. Here, Amplitude regularization is applied to the real-imaged inputs to the discriminator, which forces the generator to adapt to produce outputs that mimic data in this form.  Amplitude regularization works to encode real classical data, however given quantum states can assume complex values, we would like to ensure our generator gains access to this space. Improving the expressibility of our generator to generate complex values is the reason for Unitary injection.




\subsection{Unitary Injection}

Recall amplitude embedding embeds classical data on quantum circuits as real-valued amplitudes. While this is an efficient scheme for encoding data, we would like our generator to mimic a richer space that encompasses the complex plane, which could be used to enrich the generator. However, as the discriminator would learn to differentiate real images from synthetic ones, an easy way to differentiate would be to observe that the embedded real images do not contain any complex phases. In this case, the optimal behavior for the generator would be to avoid the complex plane altogether, which could be limiting to the space in which the quantum state can express. 

We choose to therefore inject a small, static unitary quantum network after the amplitude embedding circuit, so that the generator will be constrained to access the complex plane.  We add a unitary circuit with randomized rotation gates and entanglements, in order for the data to be represented with complex phases. Doing this allows the generator reasonably fool the discriminator that sees complex data with complex data it generates. 

Additionally, we add an inverse circuit in inference to undo this transformation that would have jumbled up the classical image. Leveraging the reversibility property of unitary operations, the inverse of a circuit $U$ is simply the reverse of the circuit. More formally, the complex conjugate transpose $U^\dag$ of any unitary $U$ is by definition $I=UU^\dag$. We apply $U^\dag$ only in inference to obtain the classical circuit. The static sub-circuit added is comprised of a few small entangling layers using uniformly distributed rotations sampled from [0,$\pi$] in addition to entanglements through CX gates. 


\vspace{2mm}

\begin{figure}[t]
    \centering
    \includegraphics[scale=0.24]{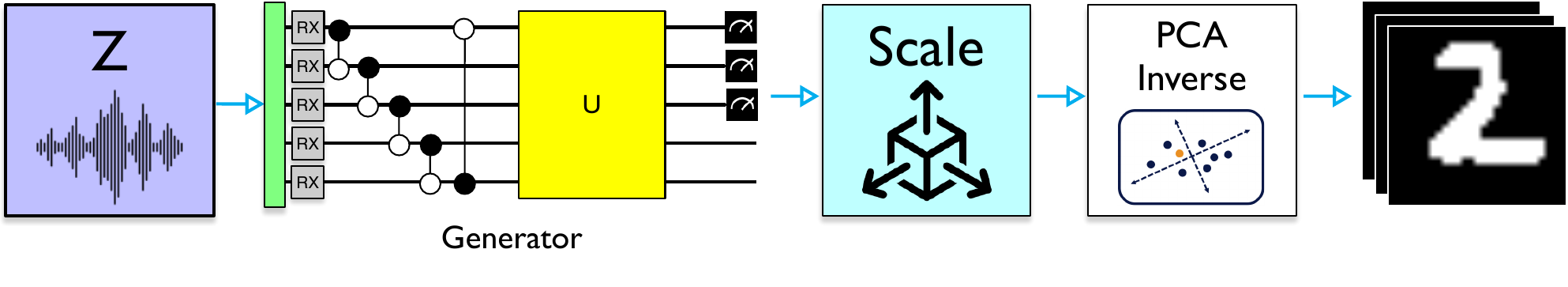}
    \caption{Inference procedure for \sol{}. Noise is sent through the quantum generator, then scaled to the domain of PCA components, before PCA inverse is performed using the eigenvalues of the original dataset.}
    \label{fig:Inference}
    \vspace{-2mm}
\end{figure}

\noindent\textbf{Inference Procedure.} The inference procedure appends the inverse of the unitary matrix that was injected into the discriminator, into the generator. In order to interpret the output classically, we apply this unitary in reverse, then take the square root of the probabilities, followed by an inverse scaling procedure to change the basis of measured values to corresponding points in the PCA domain. Following this, the PCA inverse procedure is applied to the scaled values and is observed. This can be seen in Figure \ref{fig:Inference}. 

\subsection{Putting Everything Together}

The training workflow for \sol{} is summarized in Figure \ref{fig:Passes}. Figure \ref{fig:Passes} outlines the steps for the cases of training the discriminator on real images, training the discriminator on generated images, and training the generator. As a prepossessing step before training, the images are regularized by amplitude regularization. Following this, the three steps mentioned are simultaneously trained at each training iteration in a minimax game. In the training case where the discriminator receives real images as inputs, see case (A) in Figure \ref{fig:Passes}, a gradient-free, randomized unitary operation is injected after amplitude embedding to ensure the discriminator only handles cases of data with complex phases. 

The testing workflow is visualized in Figure \ref{fig:Inference}, where the generator is attached to the inverse unitary matrix, then deregularized following measurement in order to produce classical data that resembles the input distribution to create high-quality images.

\section{Methodology}

\noindent\textbf{Experiment Framework.} We evaluate the performance of \sol{} using PennyLane \cite{pennylane} as our primary framework, while leveraging Qiskit \cite{Qiskit} for compilation. We wrap the PennyLane components in PyTorch classes which enable access to the well-tested PyTorch suite of loss functions, optimizers, etc. We use Python version 3.9.6, PyTorch version 2.0.1, PennyLane version 0.31., and Qiskit version 0.44.0.

\vspace{2mm}

\noindent\textbf{Datasets.} We evaluate the performance of \sol{} over 2 key datasets. We use MNIST as it presents a challenge of reasonable scale to NISQ-era quantum machines. This is why MNIST has been widely used for NISQ-era Quantum Machine Learning tasks \cite{silver2022quilt, wang2022quantumnas, kerenidis2018quantum}. Additionally, we use Fashion-MNIST, which presents slightly more of a challenge for \sol{}. Fashion-MNIST contains the same number of classes as MNIST and has images of the same size (28x28 pixels). While MNIST contains digits [0-9], Fashion-MNIST contains images such as sneakers, ankle boots, and t-shirts.

\vspace{2mm}

\noindent\textbf{Figure of Merit.} We analyze image quality primarily through the use of Fréchet Inception Distance (FID) \cite{heusel2017gans} which is a commonly used metric to assess image quality of generative networks \cite{karras2019stylebased}. The FID score measures how similar two distributions are, with higher scores indicating more distance and lower fidelity between the two datasets. The FID score is between a distribution with mean $\mu_1$ and covariance $C_1$ and a distribution of mean $\mu_2$ and covariance $C_2$ is defined as 
\begin{equation}
\text{FID}=  \norm{\mu_1-\mu_2}_2^2 + \text{tr}(C_1 + C_2 - 2(C_2C_1)^{1/2})
\end{equation}

\noindent\textbf{Experimental Parameters.} Our experiments are all trained for 500 training iterations, using a batch size of 20. We use a generator learning rate of $0.05$ and a discriminator learning rate of $0.05$. For training, we employ PyTorch's BinaryCrossEntropyLoss function to calculate loss and PyTorch's StandardGradientDescent optimizer. Our circuits use five qubits, each with three repeated layers, and measure the probabilities on the first three qubits, leading to 8 variables being read out, one for each PCA feature with the final readout corresponding to the regularization factor which is ignored. The best model during training is selected for inference in all experiments. Following training, we generate 100 images per model and compute the FID score across the images to their respective datasets. The FID score computed in this step is what is reported in our results.

\vspace{2mm}

\noindent\textbf{Competing Techniques.} We compare \sol{} to two competing techniques. We compare a hybrid design with a quantum generator and a classical discriminator, which we refer to as QCPatch \cite{PhysRevApplied.16.024051}. As the size of the model is much larger than \sol{}, we scale down the model to have a more comparable number of parameters. Specifically, we use 156 parameters in QCPatch (72 quantum down from 100) and which is closer to the 30 parameters used in \sol{}. Additionally, as QCPatch is not built to produce as large of an image as MNIST or Fashion MNIST directly, we interpolate the image using Bilinear Interpolation, which we found to yield the best results over other interpolation schemes. We use a different learning rate in the case of QCPatch as the classical discriminator is stronger in the given configuration. We keep the discriminator learning rate at $0.05$ and use a generator learning rate of $0.3$.

We also compare \sol{} against the Baseline design discussed in the design. Recall that the Baseline design is also fully quantum, but has a separate generator and discriminator, no amplitude regularization, and no unitary injection.

\vspace{2mm}

\noindent\textbf{Experiments on Real Quantum Hardware.}  We use the IBM Lagos machine to evaluate the performance of our model and compare the generated images to the other competing techniques. IBM Lagos is on the smaller end of superconducting quantum machines, consisting of 7 qubits, using the Falcon r5.11H processor. Due to resource constraints, we use the calibrated error model to calculate FID scores. 
\section{Results and Evaluation}

\begin{figure}[t]
    \centering
    \includegraphics[scale=0.5]{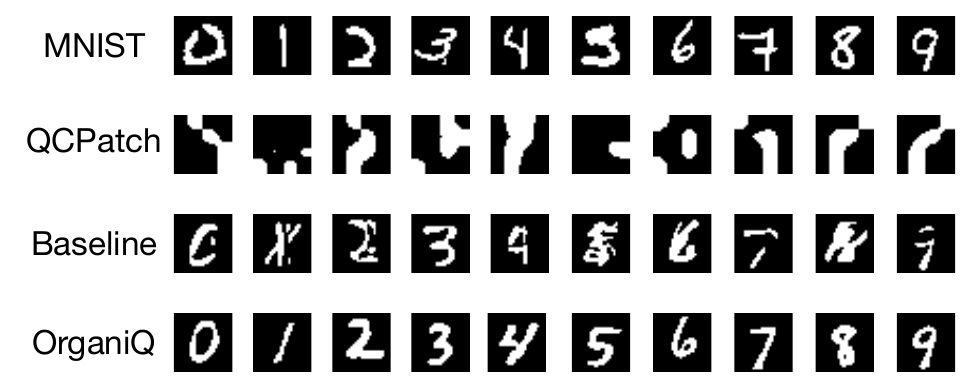}
    \caption{MNIST images generated by the Baseline, QCPatch, and \sol{}.  \sol{} generates more easily recognizable digits with less blur than other methods.}
    \label{fig:MNISTPiQ}
    \vspace{-2mm}
\end{figure}

\begin{figure}[t]
    \centering
    \includegraphics[scale=0.56]{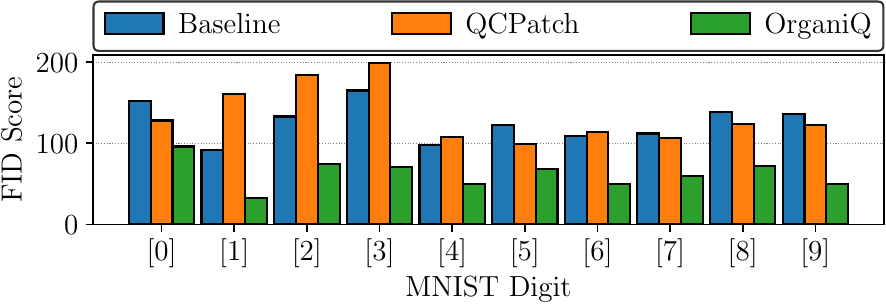}
    \caption{Image quality encapsulated as FID scores (lower is better) over the MNIST dataset. \sol{} consistently outperforms competing techniques in achieving the best image quality, despite not using any classical neural networks in training or inference.}
    \label{fig:MNIST}
    \vspace{-2mm}
\end{figure}

\begin{figure}[t]
    \centering
    \includegraphics[scale=0.5]{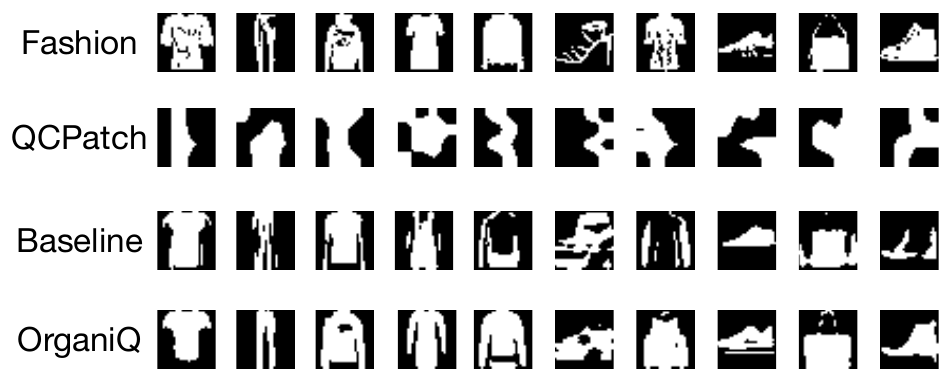}
    \caption{Fashion MNIST images by Baseline, QCPatch, and \sol{}. Fashion MNIST images tend to be more difficult for these models than MNIST.}
    \label{fig:FashionPiQ}
    \vspace{-2mm}
\end{figure}

\begin{figure}[t]
    \centering
    \includegraphics[scale=0.56]{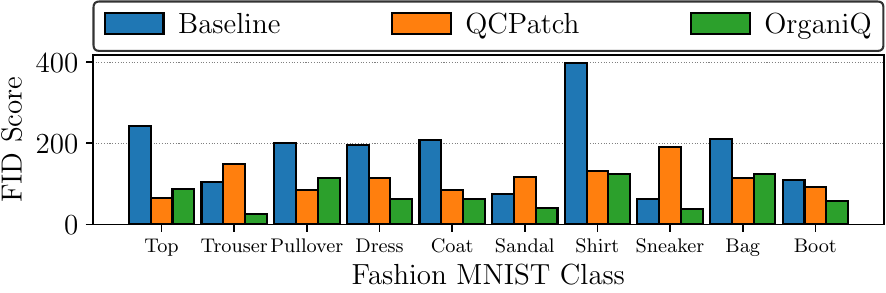}
    \caption{Image quality encapsulated as FID scores (lower is better) over the Fashion MNIST dataset. }
    \label{fig:fidfashion}
    \vspace{-2mm}
\end{figure}

\noindent\textbf{\sol{} effectively generated images across the MNIST and Fashion MNIST datasets, despite not using a classical neural network to train.} \sol{} outperforms competing hybrid techniques in all cases in the MNIST dataset as shown in Figure \ref{fig:MNIST}. 
Additionally, \sol{} significantly outperforms the Baseline approach that maintains a quantum generator and quantum discriminator in every MNIST class.
We additionally provide visual display of the generated MNIST images for \sol{} in Figure \ref{fig:MNISTPiQ}. \sol{} provides consistently high-quality images. Compared to the Baseline approach, the MNIST digits generated by \sol{} are less blurry in all cases. Additionally, the images are far more recognizable and higher-quality than QCPatch.

We find Fashion MNIST to be a more difficult dataset for image generation on the models we tested, which is a common trend given it is generally a more difficult dataset \cite{xiao2017fashion}. We also find \sol{} produces higher FID scores on Fashion MNIST than the MNIST dataset, which is unsurprising given the increased difficulty compared to MNIST. We show the images produced for \sol{}, Baseline, and QCPatch for the Fashion MNIST dataset in Figure \ref{fig:FashionPiQ} and additionally show the corresponding FID scores in Figure \ref{fig:fidfashion}.

We acknowledge that while FID scores are lower in certain cases for QCPatch, the images appear of worse quality. A feature of FID score is that it aggregates metrics regarding diversity and quality \cite{borji2022pros}; however, this can lead to cases where FID score appears low, even with poor visual quality as the diversity of images may be more in line with the original distribution. We believe this is the case with QCPatch compared to \sol{} and Baseline.

\subsection{Ablation Analysis}
\begin{figure}[t]
    \centering
    \includegraphics[scale=0.56]{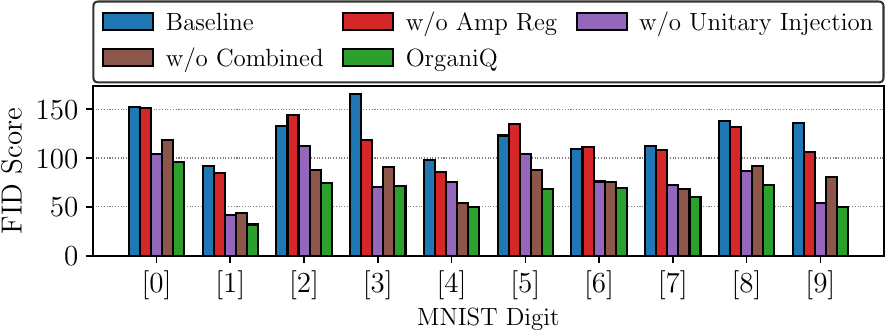}
    \caption{Ablation for the three major techniques in \sol{} for MNIST shown.  Amplitude embedding provides the largest benefit, where combined networks is the second most effective.}
    \label{fig:ablation1}
\end{figure}

\begin{figure}[t]
    \centering
    \includegraphics[scale=0.56]{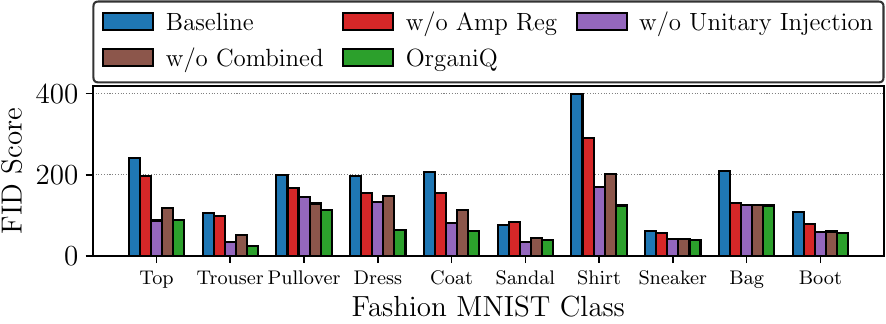}
    \caption{Ablation for the three major techniques in \sol{} for Fashion MNIST shown.}
    \label{fig:ablation2}
    \vspace{-4mm}
\end{figure}

\textbf{We perform an ablation analysis and find each of \sol{}'s techniques provide benefits to image quality} We compare the image quality results of \sol{} with the Baseline and with QCPatch to analyze how our techniques each contribute to the overall improvements in image quality. We show the corresponding FID scores without the three main techniques of connected architecture, amplitude regularization, and unitary injection over the MNIST and Fashion MNIST datasets respectfully in Figure \ref{fig:ablation1} and Figure \ref{fig:ablation2}. We find that the amplitude regularization tends to account for a large improvement in overall image quality as removing it tends to produce images of worst quality compared to removing the other techniques. In all cases between MNIST and Fashion MNIST, we find that amplitude regularization accounts for most of the improvement of \sol{} over the Baseline method. However, we still find the other techniques of unitary injection and combined networks consistently provide an improvement in overall image quality. In MNIST and Fashion MNIST, we find that the second most effective technique is the combined networks. In all cases, when the Baseline performs well, the relative improvements between techniques is more aligned.

\begin{figure}[t]
    \centering
    \includegraphics[scale=0.5]{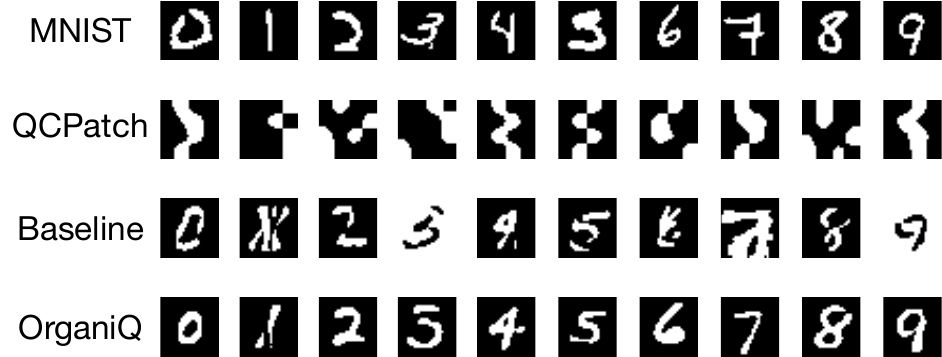}
    \caption{MNIST images generated on IBM Lagos. \sol{} creates high-quality images resembling MNIST data.}
    \label{fig:realmachinemnist}
    \vspace{-4mm}
\end{figure}

\begin{figure}[t]
    \centering
    \includegraphics[scale=0.5]{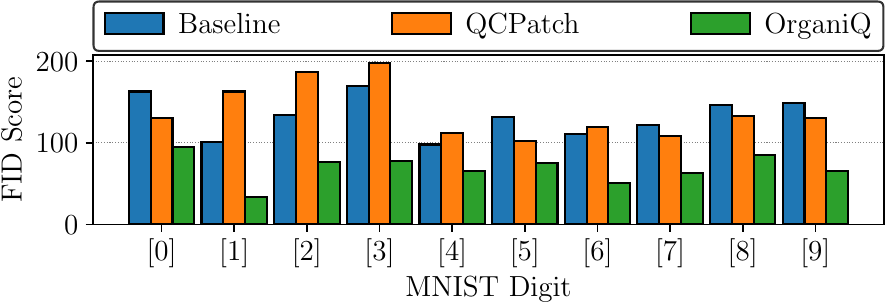}
    \caption{MNIST simulation with IBM Lagos error model.}
    \label{fig:simulationrealmnist}
    \vspace{-4mm}
\end{figure}

\begin{figure}[t!]
    \centering
    \includegraphics[scale=0.5]{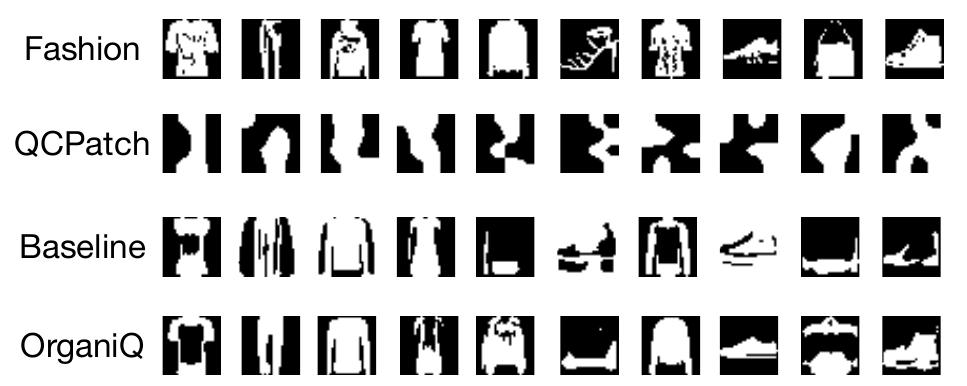}
    \caption{Fashion MNIST images generated on IBM Lagos.}
    \label{fig:realmachinefashion}
\end{figure}

\begin{figure}[t!]
    \centering
    \includegraphics[scale=0.5]{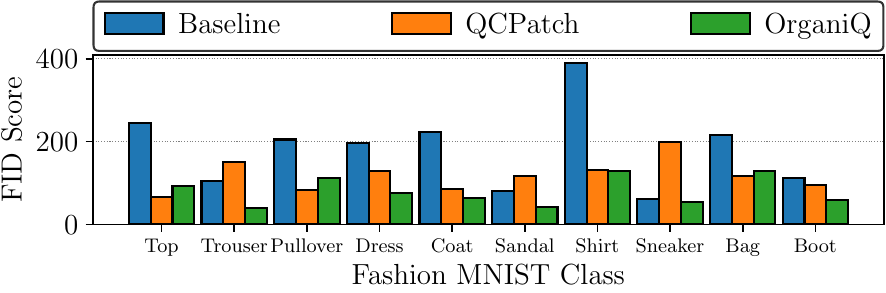}
    \caption{Fashion MNIST simulation with Lagos error model. We find \sol{} tends to outperform QCPatch in most cases and has significant quality improvements over Baseline.}
    \label{fig:simulationrealfashion}
    \vspace{-4mm}
\end{figure}

\subsection{Results on Real Quantum Machines}

\textbf{We evaluate \sol{} on a real quantum machine and find it is able to produce high quality images.} We test the feasibility on today's NISQ-era quantum machines by evaluating \sol{}, Baseline, and QCPatch on IBM Lagos quantum computer on the MNIST dataset. We provide the MNIST images generated by Lagos in Figure \ref{fig:realmachinemnist}. We find that \sol{} outperforms the other methods visually on the MNIST dataset, particularly performing well even in cases where QCPatch and Baseline struggle such as the digit 3. Additionally, we simulate image generation, using a noisy simulation model that models IBM Lagos machine and visualize the results in Figure \ref{fig:simulationrealmnist}. We find slight decreases in FID score across all three methods, with \sol{} continuing to outperform other methods on the MNIST dataset.

We evaluate \sol{} on Fashion MNIST using IBMs Lagos quantum computer. We provide the generated images in Figure \ref{fig:realmachinefashion} and find that \sol{} consistently produces high quality images over most classes in the Fashion MNIST dataset. We find the difficult cases (class 5 and class 8) which are difficult for \sol{} are likewise difficult for Baseline and QCPatch. In other cases, such as in classes 1, 2, and 7, we find \sol{} produces high quality images that are more clear than the other methods tested. We also simulate the models using IBM Lagos's error model in Figure \ref{fig:simulationrealfashion} and find slightly higher FID scores than in noise-free simulation.

\section{Concluding Discussion}

In this work, we introduce \sol{}, the first quantum GAN consisting of both a quantum generator and quantum discriminator that generates high-quality images. We demonstrate the advantages of using a connected model, amplitude regularization, and unitary injection, and training a combined model for a quantum GAN. \sol{} produces higher quality images on a real quantum machine over the state-of-the-art, in certain cases outperforming by over 100 points improvement in FID score.
\clearpage

\bibliography{main}
\end{document}